\documentclass[twocolumn]{revtex4-2}

\usepackage{graphicx}
\usepackage{amsmath,amssymb}
\usepackage{physics,siunitx}
\usepackage{pgfplotstable}
\usepackage{booktabs}
\usepackage{graphicx}
\usepackage{float}
\usepackage[english]{babel}

\newcommand{\R}{\mathrm{R}}
\renewcommand{\L}{\mathrm{L}}
\renewcommand{\a}{\hat{a}}
\newcommand{\ad}{\hat{a}^\dagger}
\newcommand{\x}{\hat{x}}
\newcommand{\xd}{\hat{x}^\dagger}
\newcommand{\MP}{\mathrm{MP}}
\newcommand{\ii}{\mathrm{i}\mkern1mu}

\renewcommand{\k}{\mathbf{k}}
\usepackage[normalem]{ulem}
\renewcommand{\H}{\hat{H}}

\begin{document}

\title{Protecting Intercavity Polaritons in Strongly Coupled Cavities}
\author{Rodrigo S{\'a}nchez-Mart{\'i}nez}
\author{Yesenia A. García Jomaso}
\author{David Ley Domínguez}
\author{César L. Ordóñez-Romero}
\author{Hugo A. Lara-García}
\author{Giuseppe Pirruccio}
\email{pirruccio@fisica.unam.mx}
\author{Arturo Camacho-Guardian}
\email{acamacho@fisica.unam.mx}
\affiliation{ Instituto de F\'isica, Universidad Nacional Aut\'onoma de M\'exico, Apartado Postal 20-364, Ciudad de M\'exico C.P. 01000, Mexico\looseness=-1}

\date{\today}

\begin{abstract}


We theoretically designed and experimentally demonstrated a mechanism to protect exotic mixed light-matter states, known as intercavity exciton-polaritons, by engineering strongly coupled optical cavities. These polaritons, shared across the coupled cavity array, exhibit remarkable robustness by preserving their intercavity nature over a wide momentum range, without compromising photon-exciton mixing or the spatial segregation of their photonic and excitonic components, which also enables a tunable heavy mass. Additionally, we unveil a direct connection between the transparency window, characteristic of slow-light experiments, and the protection of the intercavity polariton nature. Both phenomena originate from the strategic design of an energy-level landscape featuring a $\Lambda$-scheme, opening new avenues for exploring and utilizing these unique optical excitations in advanced photonic applications.

\end{abstract}

\maketitle

{\it Introduction.-}
Driven-dissipative systems that combine gain and loss elements in a single resonator have demonstrated a rich linear and non-linear phenomenology.
While linear gain-loss systems are excellent platforms to study exceptional points~\cite{Dembowski2001,Peng2016,Miri2019}, PT-symmetric~\cite{Ruter2010,Ozdemir2019,Velazquez2024}, and non-Hermitian Hamiltonians~\cite{ElGanainy2018,Ashida2020}, their nonlinear counterparts exhibit fascinating phenomena ranging from self- and parametric-oscillations~\cite{Kuznetsov2020,Savvidis2000}, self-trapping~\cite{Raghavan1999}, self-selection~\cite{Yao2023}, spontaneous symmetry breaking~\cite{Garbin2022}, and superfluidity~\cite{Keijsers2024}. 
Polaritons, hybrid half-light half-matter quasiparticles, are a typical example of driven-dissipative systems that can be operated in both linear~\cite{YuenZhou2024,SchwennickeArXiV} and non-linear regimes~\cite{Gao2015,Abbarchi2013,Savvidis2003}.
They typically emerge from the hybridization of confined photons and matter excitations that spatially overlap, such as in conventional cavities~\cite{Weisbuch1992,Carusotto2013,Basov2020}.
Polaritons have enabled the realization of quantum fluids of light ~\cite{Claude2022,Claude2023}, an exotic phase of light and matter that exhibits intriguing phenomena such as Bose-Einstein condensation~\cite{Imamoglu1996,Deng2002,Deng2003,Schneider2013}, superfluidity~\cite{Amo2009,Lerario2017}, quantum vortices ~\cite{Lagoudakis2003,Tosi2012,Panico2023}, optical bistability~\cite{Amthor2015}, out-of-equilibrium phase transitions~\cite{Kasprzak2006}, Josephson oscillations~\cite{Lagoudakis2010,Abbarchi2013}, among others.
A way to add complexity to polariton systems is by means of photonic lattices that confine light in periodic potentials ~\cite{Amo2016}. This strategy opened new opportunities to control the energy dispersion of polaritons~\cite{Wurdack2023}, unveiling topological phases of polaritons~\cite{RojasSanchez2023} and flatbands~\cite{Harder2020}.

Strongly coupled cavities offer a different and unique opportunity for the efficient and scalable generation of exotic phases of polaritons~\cite{Sawicki2021,Abbarchi2013,Sarchi2008,Rodriguez2016}, and photons~\cite{BalthasarMueller2017,Peng2014}.
Unlike conventional single cavities, cavity arrays enable the realization of a new class of polaritons, generically termed intercavity, whose field is spread across the individual coherently-coupled cavities~\cite{Sciesiek2020,Avramenko2022}.
However, when the photonic and excitonic degrees of freedom are spatially separated, a {\it pure} intercavity polariton forms with unique dynamics, band dispersion, and photoluminescence properties~\cite{GarcaJomaso2024}.
The latter class of polaritons challenges the intuitive picture of strong light-matter coupling based on the common geometry of excitons and photons overlying in the same space, replacing it with photon tunneling that couples together the light and matter degrees of freedom, remotely.

Intercavity polaritons arise as a consequence of an underlying three-level energy scheme.
Light-matter systems based on $\Lambda$-schemes have unveiled new classes of polaritons, ranging from dark-state polaritons in quantum gases~\cite{Fleischhauer2000} to dipolaritons in two-dimensional materials~\cite{Togan2018,Rosenberg2018,Datta2022,Esben2024}. In atomic gases, similar three-level $\Lambda$-schemes have been employed to observe electromagnetically induced transparency (EIT)~\cite{Harris1997,Fleischhauer2005} and slow-light~\cite{Hau1999}, where light propagates as an undamped dark-state polariton in an otherwise opaque medium. 
EIT resonances are not restricted to atomic systems, but emerge in other physical disciplines such as plasmonics~\cite{Liu2009,Okamoto2016}, magnonics~\cite{Mouadili2017,Munir2023} and it has been applied also to superconducting circuits~\cite{Andersson2020}. This demonstrates the versatility of this phenomenon across fundamentally different systems.


In this Letter, we experimentally demonstrate the robustness of the energy dispersion of pure intercavity polaritons and explain it by drawing an analogy to the transparency window observed in slow-light systems, acting as a protecting mechanism against cavity detunings.
One of the central points of the paper is the possibility to preserve the pure intercavity nature of polaritons over a wide momentum range without compromising the degree of photon-exciton mixing, nor their spatial segregation.
Our theoretical approach, based on the Green's function formalism, shows excellent agreement with experimental observations, enabling us to characterize the nature of the polaritons.  
The room-temperature operation, versatility, tunability, and scalability of our experimental setup open new avenues for generating novel forms of light and slow light that are robust over a wide angular range, enhancing its potential for the realization of new quantum many-body phases and applications.

\begin{figure}[h!]
		\centering
		\includegraphics{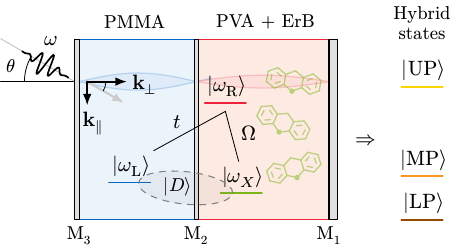}
		\caption{Schematic representation of the system and $\Lambda$-scheme.
        Two cavities are separated by a thin Ag mirror, $\mathrm{M_2}$, which allows photons to tunnel between the cavities. The left cavity is filled with PMMA, while the right cavity contains an active medium that supports Frenkel excitons.
        This configuration gives rise to a three-level $\Lambda$-scheme, where photons in the left cavity $|\omega_{\text{L}}\rangle$ tunnel to the right cavity $|\omega_{\text{R}}\rangle$ with an amplitude probability $t$.
        The light-matter coupling in the right cavity is characterized by $\Omega$. On resonance, when $\omega_\mathrm{L}(0) = \omega_X$, the middle polariton becomes a pure intercavity polariton, $|\text{MP}\rangle=\ket{D}$.
		}
		\label{fig:system}	
    \end{figure}

  	{\it Model.-} We experimentally design a system consisting of two strongly coupled cavities, as illustrated in Fig.~\ref{fig:system}. The left photonic cavity is filled with a transparent polymer, polymethylmethacrylate (PMMA), while the right excitonic cavity embeds a layer of polyvinyl alcohol (PVA) doped with a high concentration of Erythrosin B molecules (EryB). These molecules exhibit a Frenkel exciton at an energy of ${\omega_X \sim 2.24\,\text{eV}}$ that couples to the right cavity photons with an oscillatory strength denoted by $\Omega$. The cavities are separated by a thin Ag mirror, which allows photons to tunnel between them with a hopping rate $t$.

    The system can be described by the Hamiltonian 
    \begin{align}
	\label{eq:Heff}
		\nonumber
		\H =
		&
		\sum_{\k}
		\Big[
			\omega_\L(\k)\,\ad_{\L,\k}\a_{\L,\k} + 
			\omega_\R(\k)\,\ad_{\R,\k}\a_{\L,\k} +
			\omega_X \xd_{\k}\x_{\k}
		\\[-1mm]
		&
		-t\big(\ad_{\R,\k}\a_{\L,\k} + \ad_{\L,\k}\a_{\R,\k} \big)
			+\Omega\big(\ad_{\R,\k}\x_\k + \xd_\k\a_{\R,\k}\big)	
		\Big],
	\end{align}
    the first line represents the kinetic energy of photons ($\a_{\L/\R,\k}$) and excitons ($\x_\k$). The first term in the second line describes the tunneling of photons between cavities, while the second term accounts for the light-matter coupling. Here, $\mathbf{k}$ denotes the in-plane momentum.
    Experimentally, $\mathbf{k}$ is controlled by the angle of incidence, $\theta$ (see Fig.~\ref{fig:system}), because $|\k| = (n^\text{eff}\omega_0/c)\sin\theta$, where $\omega_0 = c\pi/(n^\text{eff}L)$ is the frequency of light at normal incidence, and $L$ and $n^\text{eff}$ are the cavity width and the effective refractive index, respectively. The dispersion of the cavity photons in terms of the angle of incidence is given by \cite{Kavokin2017microcavities} as
    \begin{equation}
    	\omega_i(\theta)
    	= \frac{\omega_{i,0}}{\sqrt{1-(\sin\theta/n_i^\text{eff})^2}},
    \end{equation}
    and we assume infinite-mass excitons such that their dispersion $\omega_X(\mathbf k)=\omega_X$ can be regarded as momentum independent.

    {\it Intercavity polaritons.-} The three eigenmodes of the coupled cavities give three polariton branches: lower (LP), middle (MP) and upper (UP) polariton. If the exciton energy lies on resonance with the left cavity at normal incidence, $\omega_{\L,0} = \omega_X$, the MP becomes a \textit{pure intercavity polariton}, decoupled from the right cavity mode,
    \begin{equation}\label{eq:intercavity_MP}
        \ket{D(\mathbf k=0)} = \cos\phi\ket{\omega_\L(\k = 0)} + \sin\phi\ket{\omega_X},
    \end{equation}
   with a mixing angle given by $\tan\phi = t/\Omega.$ Interestingly, the middle polariton is formed as a linear superposition of a left cavity photon and an exciton sitting in the right cavity, that is, their components are spatially segregated.
   This unique feature changes the band, underlying dynamics, photoluminiscence spectrum~\cite{GarcaJomaso2024}.

   Around normal incidence, one can obtain analytically the dispersion of the MP,
	\begin{equation}
        \label{eq:dispersion_MP}
        \omega_\MP(\k)
        \simeq \frac{\k^2}{2m_\MP} + \omega_X,
    \end{equation}
   here, the mass of the heavy polariton is given by $m_\MP/m_\L = 1 + t^2/\Omega^2 > 1$, where $m_\L = \omega_{\L,0}/(c/n_\L^\text{eff})^2$ is the mass of the left cavity photon. It is important to stress that, at normal incidence and under resonance conditions, the properties of the intercavity polariton depend solely on the hopping $t$ and the light-matter coupling $\Omega$, and are completely independent of the frequency of the right cavity mode.
    Away from normal incidence, the resonance condition no longer holds, and the pure intercavity polariton can break down. Interestingly, the right cavity frequency, $\omega_{\R,0}$, can be exploited to control the angular regime in which the middle branch remains an effective intercavity polariton.
    
\begin{figure}[h!]
        \centering
        \includegraphics{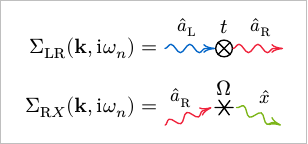}
        \caption{
           Self-energy terms.
    The wavy colored lines denote the propagators of the bare states of the system: the left and right cavity photons, $\a_{\L/\R}$, and the exciton $\x$. The circled cross and the asterisk represent the tunneling coefficient and the light-matter coupling, respectively. We have $\Sigma_{ij}=\Sigma_{ji}.$
        }
        \label{fig:Feynman-diagram}
    \end{figure}

    {\it Theoretical formalism.-} Before delving into the experimental results, let us first theoretically understand the different regimes of intercavity polariton formation as a function of in-plane momentum, i.e., beyond normal incidence.
Our theoretical approach is based on a time-imaginary Green's function formalism. To this end, we introduce the spinor
    \begin{equation}
        \hat\Psi^\dagger(\k,\tau) = [\hat a^\dagger_{\L,\k}(\tau),\hat a^\dagger_{\R,\k}(\tau),\hat x^\dagger_{\k}(\tau)],
    \end{equation}
    and the Green's function
    \begin{equation}
        \mathcal G_{ij}(\k,\tau) = 
        -\langle T_\tau\{\hat{\Psi}_i(\k,\tau)\,\hat{\Psi}^\dagger_j(\k,0)\} \rangle,
    \end{equation}
    where $i,j\in\{\L,\R,X\}$ represents the associated quantum field and $T_\tau$ indicates the time-ordering operator.
    In momentum-energy space, the Green's function obeys the Dyson equation,
    \begin{equation}
    \label{eq:Green_function}
       \mathcal G^{-1}_{ij}(\k,\ii\omega_n) = 
        G_{ij}^{-1}(\k,\ii\omega_n)
        -\Sigma_{ij}(\k,\ii\omega_n),
    \end{equation}
    with $\omega_n$ a bosonic Mastubara frequency, $\Sigma_{ij}(\k,\ii\omega_n),$ the self-energy and $G_{ij}(\k,\ii\omega_n)$ the ideal Green's function, which reads as a diagonal matrix:
    \begin{gather}
        \nonumber
        {G}_{\L\L}
        = \frac{1}{\ii\omega_n - \omega_\L(\k)}, 
        \quad
        {G}_{\R\R}
        = \frac{1}{\ii\omega_n - \omega_\R(\k)}, \\
        {G}_{XX}
        = \frac{1}{\ii\omega_n - \omega_X}.
    \end{gather}
    Here, the self-energies are depicted diagramatically in Fig.~\ref{fig:Feynman-diagram}.

    By inverting \eqref{eq:Green_function} we find
    \begin{align}
        \nonumber
        \mathcal G^{-1}_{\L\L}(\k,\ii\omega_n)
        &= \ii\omega_n - \omega_{\L}(\k) + \ii\eta_\L \\ 
        \label{eq:G_LL}
        &\quad -\frac{t^2}{\ii\omega_n- \omega_{\R}(\k) + \ii\eta_\R - \frac{\Omega^2}{\ii\omega_n - \omega_X}},
    \end{align}
    for the left cavity Green's function.   From this expression, we can straightforwardly extract the polariton branches, which emerge as poles of the Green's function.
   We have included the damping rates $\eta_{\L/\R}$ to account for photon leakage through the cavities.
   Furthermore, we can easily obtain the quasiparticle residue of the branches, which corresponds to the squared Hopfield coefficients.
   In our analysis, we focus on the MP, which, on resonance, emerges as a pole with energy $\omega_{\L,0} = \omega_X$ and a quasiparticle residue of ${ 1/(1 + \frac{t^2}{\Omega^2})}$.
   It is important to recall that the quasiparticle residue represents the composition of the left cavity photon that forms the intercavity polariton.
    
	\begin{figure}[h]
    	\centering

        \includegraphics{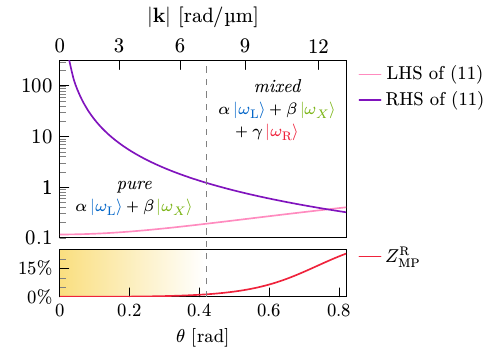}
    	\caption{
           Intercavity polaritons transition from a pure, spatially segregated state to a mixed one.
           (Top) The left-hand side (LHS) and right-hand side (RHS) of Eq.~\eqref{eq:breaking_condition}, plotted on a logarithmic scale, determine the breakdown of the pure intercavity polariton as a function of $\k$ and and the corresponding incidence angle $\theta$.
           When these terms become comparable, the pure intercavity polariton transitions into a mixture of the two bare photonic states and the exciton.
           (Bottom) The quasiparticle residue (Hopfield coefficient) quantifies the contribution of right-cavity photons to the MP state, showing that it remains a pure intercavity polariton with a vanishing residue over a wide angular range.
           The yellow shading highlights the crossover from a pure intercavity polariton to a mixed polariton state.
           The dashed line marks the point where the right-cavity photons contribute more than 1\% to the MP state.
           At this threshold, the LHS and RHS of Eq.~\eqref{eq:breaking_condition} differ by approximately one order of magnitude.
       }
        \label{fig:breaking_condition}
    \end{figure}
    
    For the MP to retain its pure intercavity nature as the momentum deviates slightly from zero, it should remain decoupled from the right cavity photon.
    From Eq.~\ref{eq:G_LL} we read that this holds for
    \begin{equation}
    \label{condition}
    |\omega_{\MP}(\k) - \omega_X|
        \ll 
         \frac{\Omega^2}{    \big|\,\omega_{\MP}(\k) - \omega_\R(\k) + \ii\eta_\R\,\big|},
    \end{equation}
    the right-hand side of this equation is the equivalent of a transparency window found in EIT and for the propagation of dark-state polaritons in quantum gases~\cite{Fleischhauer2005}. In our case, this equation is physically very intuitive: As long as the dispersion of the middle polariton remains inside the {\it transparency window} the middle polariton retains its pure intercavity character.
    
    Taking a parabolic approximation to describe the photonic, $\omega_\R(\k) = \omega_{\R,0} + \k^2/(2m_\R)$, and polaritonic,  eq.~\eqref{eq:dispersion_MP}, dispersions we re-write the condition in Eq.~\ref{condition} as
    \begin{equation}
    \label{eq:breaking_condition}
    \sqrt{\left(
        \frac{\k^2}{2\bar{m}}
        + \delta_\R
    \right)^2 + \eta_\R^2}
    \ll 
    \frac{\Omega^2}{\k^2/(2m_\MP)},
    \end{equation}
    with $1/\bar{m} = 1/m_\R - 1/m_\MP > 0$ being an effective mass and $\delta_\R := \omega_{\R,0} - \omega_X$, the right-cavity detuning. 
    It is clear that for very large $\k$, the left-hand side (LHS) of the inequality grows roughly as $\k^2$, whereas the right-hand side (RHS) diminishes proportionally to $1/\k^2$.
    This perfectly illustrates the fact that intercavity polaritons tend to break for large momentum.
    In Fig.~\ref{fig:breaking_condition} (top), we show the LHS and RHS of Eq.\eqref{eq:breaking_condition}, while in Fig.~\ref{fig:breaking_condition} (bottom), we show the right-cavity quasiparticle residue of the MP, $Z_\MP^\R = |\braket{\omega_\R}{\MP}|^2$ that determines the amount of left-cavity photon forming the MP state. 
    For normal incidence ($\mathbf{k} = 0$), the RHS diverges, emphasizing that for $\mathbf{k} = 0$, the MP is strictly a pure intercavity polariton.
    Here, the quasiparticle residue in Fig.~\ref{fig:breaking_condition} (bottom) is exactly zero.
    Slightly away from normal incidence, the RHS is orders of magnitude larger than the LHS, and the MP retains its intercavity nature; in this regime, the residue $Z_\MP^\R$ is negligible.
    As $\theta$ increases, we observe that the LHS and RHS become comparable, leading to the breaking of the pure intercavity polariton, which is also reflected in the quasiparticle residue that increases significantly.
    In this regime, the $\ket{\MP}$ state is a mixture of the three bare states.
    The transition from a pure intercavity polariton to an admixture is not sharp, instead it is a smooth transition, illustrated by the yellow shading in Fig.~\ref{fig:breaking_condition} (bottom).

   Interestingly, modifying the right-cavity properties provides a mechanism to circumvent the breaking of the intercavity polariton at large angles.
   The key element is the detuning of the right cavity, $\delta_\R$, which can be adjusted to compensate for the growth of $\k^2 / (2\bar{m})$ in the LHS of \eqref{eq:breaking_condition}. This tuning stabilizes the system and allows for the emergence of a robust, effective intercavity polariton.

    In the context of slow light in atomic gases, intriguing phases of polaritons—such as superfluidity above Landau's criterion~\cite{Nielsen2020} and the ability to use polaritons in a non-demolition scheme to probe universal features of polarons~\cite{Camacho2023}—have been predicted, exploiting the engineering of energy level schemes. Here, we demonstrate an experimental realization of a similar mechanism.

    \begin{figure}[H]
		\centering
		\includegraphics{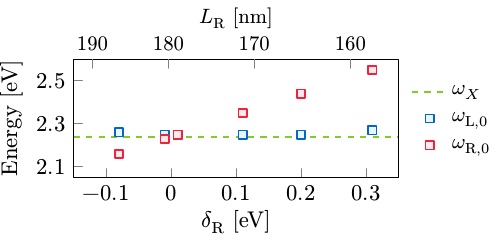}
		\caption{
    		Best fits to the cavity energies at normal incidence, $\omega_{i,0}$ of the left (blue quares) and right (red squares) cavities, for six representative measurements.
            The dashed green line represents the exciton energy and demonstrates that the resonance condition $\omega_{\L,0} = \omega_X$ is essentialy mantained across the sample.
		}
		\label{fig:trend}
	\end{figure}
    \begin{figure*}[ht]
	\centering
		\includegraphics{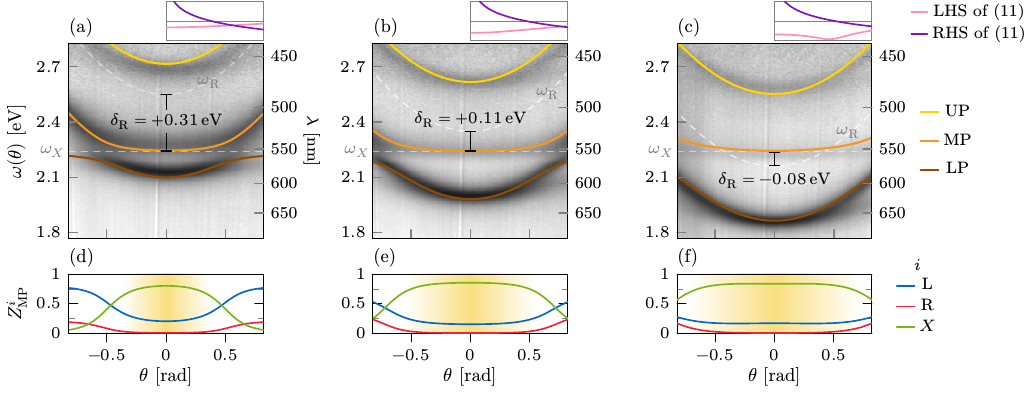}
	\caption{
		(a-c) Reflectance spectrum (in white and black) for different values of the right cavity detuning $\delta_\R$.
        The color lines correspond to the theoretical modelling of the three polariton branches.
        The white dashed line depicts the exciton, $\omega_X$, and right-cavity photon $\omega_\R(\theta)$, energies.
        (d-f) The Hopfield coefficients for the middle polariton.
        The blue/red lines correspond to the left/right cavity component to the MP while the dashed line gives the excitonic component.
        The yellow shadowing illustrates the breakdown of the pure intercavity nature of the MP giving by the condition in Eq.~\ref{eq:breaking_condition}, comparison which we illustrate as an inset on top of each reflectance spectrum.
	}
	\label{fig:selected_maps}
	\end{figure*}
          
    {\it Experimental Results.-}
    To experimentally demonstrate our theoretical predictions, we design our coupled-cavity so that it satisfies the resonance condition $\omega_{\L,0} = \omega_X$, while allowing for a wide range of tunable right-cavity energies.
    In Fig.~\ref{fig:trend}, we plot the cavity energy values for normal incidence and for six different points on the sample, determined by performing numerical fits to the corresponding reflectance spectra.
    The blue squares represent the energy of the left cavity, showing that it remains essentially resonant with the exciton energy (green dashed line) for all of the measurements we performed. The tunability of $\omega_{\R,0}$ (red squares) through adjustment of the cavity thickness, $L_\R$, allows us to tailor the dispersion of the pure intercavity polariton through the detuning parameter $\delta_\R$, as explained in the previous section.

    Our main results are shown in Fig.~\ref{fig:selected_maps} where we display the reflectance spectra of the sample for different values of the right-cavity detuning (a-c), together with the quasiparticle residues (d-f), which describe the composition of the MP in terms of the bare states.
    As an inset at the top, we display the LHS and RHS of Eq.\ref{eq:breaking_condition}, which guide the yellow shading in Figs.~\ref{fig:selected_maps} (d)-(f), marking the breakdown of the pure intercavity polariton.
    
    For a positive detuning, as shown in Fig.~\ref{fig:selected_maps} (a) (${\delta_\R = +0.31\,\text{eV}}$), we show that the intercavity polariton appears as a heavy polariton with suppressed dispersion around normal incidence.
    However, away from normal incidence, the polariton rapidly becomes dispersive, indicating the breakdown of the pure intercavity polariton state.
    Indeed, in Fig.~\ref{fig:selected_maps} (d), we observe that when the MP becomes dispersive, the residue of the right cavity photon increases, marking the formation of a polariton that is a mixture of the three bare states.
    At the top, the inset shows that the breakdown of the pure intercavity polariton coincides with the energy condition explained in Eq.~\ref{eq:breaking_condition}.

    Decreasing the right-cavity detuning, for ${\delta_\R = +0.11\,\text{eV}}$, in Fig.~\ref{fig:selected_maps} (b), we observe that the angular region for which the heavy-mass polariton persists widens.
    Correspondingly, the residues in Fig.\ref{fig:selected_maps} (e) illustrate that the MP retains its pure intercavity nature over the same angular region.
    Again, we can explain this behavior from the interplay between the terms in Eq.~\ref{eq:breaking_condition}, as shown in the inset.

    The ability to protect intercavity polaritons is further illustrated in Fig.~\ref{fig:breaking_condition} (c), which corresponds to the negative detuning, ${\delta_\R = -0.08\,\text{eV}}$.
    In this case, we observe that the MP becomes quasi-flat.
    That is, varying the detuning of the right-cavity to negative values serves as a mechanism to preserve the intercavity polariton.
    Here, the quasiparticle residue of the right cavity photon essentially vanishes, even for large angles.
    Note that for negative detunings, the LHS of Eq.\ref{eq:breaking_condition} (in pink) becomes a non-monotonic function of $\theta$.

{\it Conclusions and Outlook.-}
    Engineering the properties of exotic phases of light and matter on demand is key to enhance their complexity and scalability.
    In this Letter, we experimentally and theoretically demonstrate the control of intercavity polaritons.
    By designing a system of strongly coupled cavities, we realized pure intercavity polaritons with spatially segregated photonic and excitonic components.
    Using a Green's function formalism, we explain that upon cavity detuning, the intercavity nature of the polariton is preserved over a wide angular region. Due to their intrinsic three-level scheme, intercavity polaritons are analogous to slow light in atomic physics. In both systems, it is possible to highly control the dispersion of light. In our case, slow-light or heavy-mass polaritons arise from the interplay between photon tunneling and light-matter interactions, features that can be easily tuned in the experiment.

    Our work opens the door to studying strongly correlated phases of delocalized polaritons. In particular, we anticipate that the ability to produce heavy polaritons over a wide angular regime will enhance the effects of polariton-polariton interactions, promoting quantum many-body phases such as the Bose-Einstein condensation of intercavity polaritons, bistabilities, and highly nonlinear effects. Moreover, the unique spatial separation of the light and matter degrees of freedom observed across several detunings foster the possibility to perform position-dependent quantum tomography, as well as performing advanced experiments where charge transport and light emission are monitored simultaneously in two separated yet coupled regions of the sample. From the standpoint of novel applications, room-temperature coupled polariton lasers may be operated by electrical injection in one cavity while monitoring light emission from the other. Our architecture may contribute to solving some of the issues related to the photonic integration of polaritonic devices in optoelectronic circuits\cite{Wang2015}. Finally, we envisage the possibility of including an in-plane photonic lattice combined with a vertically stacked cavity array to achieve three-dimensional topologically protected states or to extend the slow-light analogy for periodic systems.

{\it Acknowledgments.-} The authors acknowledge financial support from the CONAHCYT-Mexico Synergy project 1564464. G.P. acknowledges Grant UNAM DGAPA PAPIIT under Nos. IN104522 and IN104325, CONAHCyT Project Nos. 1564464 and 1098652 and project PIIF 24. H.A.L-G acknowledges financial support from UNAM DGAPA-PAPIIT Grant No. IN106725 and IA107023. The work of Y.A.G J. was supported by the UNAM Posdoctoral Program (POSDOC). R.S.M acknowledges support from CONAHCyT PhD scolarship. C.L.O.R acknowledges financial support from UNAM DGAPA PAPIIT Grant No. IG101424. A.C-G. acknowledges financial support from UNAM DGAPA PAPIIT Grant No. IA101923 and IA101325,  UNAM DGAPA PAPIME Grants No. PE101223 and No. PIIF23 and Project CONAHCYT No. CBF2023-2024-1765.
\bibliography{refs_two_cavity} 

\end{document}